\documentclass[11pt]{article}

\usepackage[margin=2.5cm]{geometry}
\usepackage{amsthm}
\usepackage{amsmath}

\theoremstyle{definition}
\newtheorem{dfn}{Definition}[section]
\theoremstyle{remark}

\theoremstyle{plain}

\newtheorem{prop}{Proposition}[section]

\newcommand{\calr}{\mathcal{R}}

\newcommand{\R}{\textbf{R}}

\begin{document}
\title{Galilean relativity and its invariant bilinear forms}
\author{H.M. Ratsimbarison\\
       Institut @-HEP, Antananarivo}
\date{April, 2006\footnote{Typos corrected and text clarity improved: February 19, 2024.} }

\maketitle
\begin{abstract}
We construct the family of bilinear forms g$_G$ on $\R^{3+1}$ for which Galilean boosts and spatial rotations are isometries. The key feature of these bilinear forms is that they are parametrized by a Galilean invariant vector whose physical interpretation is rather unclear. Towards the end of the paper, we construct the Poisson bracket associated with the (nondegenerate) antisymmetric part of g$_G$.
\end{abstract}

\section{Introduction}
Due to Einstein, we know that Maxwell's equations imply the invariance of the light speed c under inertial transformations (Lorentz transformations). Mathematically, this result can be deduced from the invariance of some inner product on the spacetime under Lorentz transformations.\\
In this paper, we give all bilinear forms g$_G$ on $\R^{3+1}$ which admit Galilean transformations as inversible isometries. These bilinear forms are parametrized by a vector ($\vec{a}$,a$^0$) of $\R^{3+1}$ which is Galilean invariant. We then construct the Poisson bracket associated with the (nondegenerate) antisymmetric part of g$_G$.

\section{Galilean bilinear forms}
At the outset, we will provide all necessary definitions to avoid ambiguities. All vector spaces considered in this paper are finite-dimensional real vector spaces, unless otherwise stated.
\begin{dfn} Let (M,g) be a vector space M equipped with a bilinear form g, then a linear operator A is an \emph{isometry} of (M,g) iff it preserves g, i.e. g(Ax,Ay) = g(x,y) for all vectors x,y in M.
\end{dfn}
\begin{prop} For symmetric g, A is an isometry of (M,g) iff  g(Ax,Ax) = g(x,x) $\forall x\in M$.
\end{prop}
\textsl{Proof}: The necessary condition is trivial whereas the sufficient condition comes from :
\begin{eqnarray*}
	g(A(x+y),A(x+y)) &=& g(Ax,Ax) + g(Ay,Ay) + 2g(Ax,Ay) \\
	&=& g(x,x) + g(y,y) + 2g(Ax,Ay)\\
\textrm{In the other hand, }g(A(x+y),A(x+y)) &=& g(x+y,x+y) = g(x,x) + g(y,y) + 2g(x,y).
\end{eqnarray*}
Q.E.D.\\

When g is symmetric nondegenerate \footnote{g is nondegenerate iff g(x,y) = 0 $\forall$y$\in$M $\Rightarrow$ x = 0.}, the definition of (inversible) isometry can be provided by the notion of orthogonal operator. For this, we need to define an involution on operators of (M,g). 
\begin{dfn} Let g be a nondegenerate symmetric bilinear form on M, A a linear operator on M, then the \emph{involution} A* of A is defined by: 
\begin{eqnarray*}
	g(A^*x,y) := g(x,Ay) \quad \forall x,y\in M.
\end{eqnarray*}
An \emph{orthogonal} operator A on (M,g) is such that A*A = AA* = Id.
\end{dfn}
The nondegeneracy of g is necessary to define A* as application whereas the symmetry condition is to ensure the involution propriety A** = A. Now, we obtain the result:
\begin{prop} For nondegenerate symmetric g, a linear operator A on (M,g) is an isometry iff A*A = Id. 
\end{prop}
\textbf{Proof}:\\
The necessary condition comes from the relations:
\begin{eqnarray*}
	&& g(x,y) = g(Ax,Ay) = g(A^*Ax,y)  \quad \forall x,y \in M,\\
	\Rightarrow && g(A^*Ax - x,y) = 0 \quad  \forall x,y \in M,\\
	\Rightarrow && A^*A = Id, \textrm{ (by the nondegeneracy of g)}, 
\end{eqnarray*}
whereas the sufficient condition part is obvious.\\
Q.E.D. \\

To construct g explicitly, we can fix a basis on M and find the matrix representation of g in this basis. In a fixed basis of M, let's write:
\begin{eqnarray*}
  g(x,y) &=& g_{\mu\nu}x^{\mu}y^{\nu} , \quad g(Ax,Ay) = g_{\rho\sigma}A^{\rho}_{\mu}x^{\mu}A^{\sigma}_{\nu}y^{\nu} \quad \forall x,y \in M. 
\end{eqnarray*}
So a linear operator A on (M,g) is an isometry iff we have the following matrix relation:
\begin{eqnarray}
	g_{\mu\nu} = g_{\rho\sigma}A^{\rho}_{\mu}A^{\sigma}_{\nu}. 
	\label{key}
\end{eqnarray}

Now, we begin the construction of Galilean relativity with the definition of a Galilean transformation, which preserves  Galileo's law of inertia for inertial frames in Classical Mechanics.
\begin{dfn}  Let $\calr$ be an inertial frame, then a \emph{Galilean boost} $\Lambda_{\vec{v}}$ with velocity $\vec{v}$ on $(\textbf{R}^{3+1},\calr)$ is a linear transformation defined on Cartesian coordinates by:
\begin{eqnarray*}
	\Lambda_{\vec{v}} : (\R^{3+1},\calr) \ni (\vec{x},t) \mapsto (\vec{x} + \vec{v}t,t) \in (\R^{3+1},\Lambda_{\vec{v}}\calr),
\end{eqnarray*}
where $\Lambda_{\vec{v}}\calr$ is the inertial frame which coincides with $\calr$ at time 0, and moving with constant speed $\vec{v}$ with rapport to $\calr$.
\end{dfn}
Similarly to Special Relativity, a 4-vector ($\vec{x}$,t) in $(\R^{3+1},\calr)$ can be viewed as event seen at the position $\vec{x}$ in $\calr$, at time t. In Galilean relativity, time is not affected by Galilean boosts, so there is an \emph{absolute} time for all inertial frames.

\paragraph{Remark:} Galilean boosts are no longer \emph{linear} transformations when we reformulate them in a pure spatial space. It is then natural to use the notion of spacetime for both Special and Galilean relativities.

\paragraph{Problem:} Construct all (Galilean) bilinear forms $g_G$ on $\textbf{R}^{3+1}$ for which all Galilean transformations on ($\R^{3+1}$,g$_G$) are isometries.\\

For simplicity, we will give the construction of g$_G$ in $\R^{2+1}$ and show that the solution on $\R^{3+1}$ comes naturally from those on $\R^{2+1}$.\\
Let's consider an arbitrary Galilean boost $\Lambda_{\vec{v}}$ on $\R^{2+1}$ with velocity $\vec{v}$ = (v$^1$,v$^2$), then
\begin{eqnarray*}
 && \Lambda_{\vec{v}}
	\begin{pmatrix}
 x^1 \\
 x^2\\
 t
\end{pmatrix}
=
\begin{pmatrix}
 1 &0 & v^1 \\
 0 &1 & v^2 \\
 0 &0 & 1
\end{pmatrix}
\begin{pmatrix}
 x^1\\
 x^2\\
 t
\end{pmatrix} \quad \forall x \in \R^{2+1},\\
\textrm{and} && g_{\mu\nu} = g_{\rho\sigma}(\Lambda_{\vec{v}})^{\rho}_{\mu}(\Lambda_{\vec{v}})^{\sigma}_{\nu} \quad  \forall \vec{v} \in \R^2\\
\Leftrightarrow && 
\begin{pmatrix}
 \begin{pmatrix}
   g_{11} & g_{21} & g_{31}
 \end{pmatrix}
 & \begin{pmatrix}
   g_{12} & g_{22} & g_{32}
\end{pmatrix} 
 & \begin{pmatrix}
   g_{13} & g_{23} & g_{33}
   \end{pmatrix}
\end{pmatrix}
\begin{pmatrix}
 1 &0 & v^1 \\
 0 &1 & v^2 \\
 0 &0 & 1
\end{pmatrix}
\begin{pmatrix}
 1 &0 & v^1 \\
 0 &1 & v^2 \\
 0 &0 & 1
\end{pmatrix}\\
&& =
\begin{pmatrix}
 \begin{pmatrix}
   g_{11} & g_{21} & g_{31}
 \end{pmatrix}
 & \begin{pmatrix}
   g_{12} & g_{22} & g_{32}
\end{pmatrix} 
 & \begin{pmatrix}
   g_{13} & g_{23} & g_{33}
  \end{pmatrix}
\end{pmatrix}
\quad \forall \vec{v} \in \R^2,\\
\Leftrightarrow &&\left\{ 
\begin{aligned}
  &	g_{11} = g_{12} = g_{21} = g_{22} = 0,\\
  & g_{13} + g_{31} = g_{23} + g_{32} = 0
\end{aligned}
  \right.\\
\Leftrightarrow && g_{\mu\nu} = 
\begin{pmatrix}
 \begin{pmatrix}
   0 & 0 & - g_{13}
 \end{pmatrix}
 & \begin{pmatrix}
   0 & 0 & - g_{23}
\end{pmatrix} 
 & \begin{pmatrix}
   g_{13} & g_{23} & g_{33}
  \end{pmatrix}
\end{pmatrix}.
\end{eqnarray*}
So the general form of g$_G$ is given by:
\begin{eqnarray*}
	(g_{G})_{\mu\nu}x^{\mu}x'^{\nu} = g_{13}x^1t' - g_{13}tx'^1 + g_{23}x^2t' - g_{23}tx'^2 + g_{33}tt' \quad \forall x,y \in \R^{2+1}. 
\end{eqnarray*} 
Finally, the last equality can be written as \footnote{The standard inner product on $\R^3$ will be denoted by $\cdot$.}:
\begin{eqnarray}
	g_G(x,x') = \vec{a}\cdot (t'\vec{x} - t\vec{x}') + a^0tt' \quad \forall x,x'\in \R^{2+1},
	\label{sol}
\end{eqnarray}
where ($\vec{a}$,a$^0$) is invariant under Galilean boosts (like acceleration).

\subparagraph{Remarks:}
1) The generalization of g$_G$ to higher dimensional space $\R^{n+1}$ is obvious.\\
2) g$_G$ is a nonsymmetric bilinear form so we cannot speak of orthogonal operators on ($\R^{3+1}$,g$_G$).\\
3) By its definition, $g_G$ is not invariant under spatial translations \emph{unless} it involves two 4-vectors having the same time-component. In this case, the spatial 'distance' $\vec{x} -\vec{x}'$ is invariant under translation.\\
4) g$_G$ is not also invariant under time translation.

\subparagraph{Questions:}
\begin{itemize}
  \item Physically, what does the vector a = ($\vec{a}$,a$^0$)$\in \R^{3+1}$ represent?
	\item What is the 4-vector associated with the momentum $\vec{p}$ ?
	\item What is the group of invertible isometries Ga(3+1) of g$_G$?
	\item Is there a relation between the c $\rightarrow \infty$ limit of Minkowski metric and the Galilean bilinear form?
	\item Is there a relation between the symplectic structure of classical mechanics and the antisymmetric part of $g_G$?
\end{itemize}
Now, let us answer some of the above questions.\\
As mentioned earlier, the vector ($\vec{a}$,a$^0$) is Galilean invariant, so it behaves like acceleration. Moreover, it is necessary to have a commun dimension for the 2 parts of g$_G$, i.e.
\begin{eqnarray*}
	\left[\vec{a}\right]\left[\vec{x}\right]\left[t\right] = \left[a^0\right]\left[t\right]^2 \quad \textrm{or} \quad \left[\vec{a}\right] = \left[a^0\right]\left[\vec{x}\right]^{-1}\left[t\right].
\end{eqnarray*}
At this point, we have not yet found a direct physical interpretation of the vector a.
 
\begin{prop} The 4-vector associated with $\vec{p}$ is ($\vec{p}$,m).
\end{prop}
\textsl{Proof}: From $g_G$(dx,dx) = a$^0(dt)^2$ , we conclude that dt is Galilean invariant, so v := dx/dt = ($\vec{v}$,1) is a 4-vector, and mv = ($\vec{p}$,m) =: p.\\
 Q.E.D.
 
\subparagraph{Remarks:}
1) In a Minkowski space, the 4-vector associated with $\vec{p}$ is ($\vec{p}$,E/c), where E is the energy of the system, and one derive from the metric invariance the famous equality: E$_{rest}$ = mc$^2$. In Galilean space, the analogous relation is trivial.\\
2) When two 4-vectors x,x' have the same time component, i.e. measured at the same time, then the associated spatial distance is Galilean invariant.\\
3) The set of 4-vectors with time 0 is a Galilean invariant subspace of $\R^{n+1}$ which means that all inertial frames are coinciding at time 0. 

\section{Poisson structure from Galilean bilinear forms}

Now, let's show that the antisymmetric part of g$_G$ provides a Poisson structure on the manifold $\R^{3+1}$. 
\begin{prop} The antisymmetric part $g_{G,as}$ of $g_G$ is a nondegenerate .
\end{prop}
\textsl{Proof}:\\
We have:
\begin{eqnarray*}
g_{G,as}(x,x') := \vec{a} \cdot (t'\vec{x} - t\vec{x}') = \vec{0} \quad \forall x' \Rightarrow \quad t'\vec{x} - t\vec{x'} = \vec{0} \quad \forall x'
\Rightarrow \quad x = 0.
\end{eqnarray*}
Q.E.D.
\begin{dfn} A \emph{symplectic form} on a smooth manifold X  is a closed 2-form $\omega \in \Omega^2$(X) which is nondegenerate at each point of X.
\end{dfn}

The 2-form  $\omega_G$ := $(g_{G,as})_{\mu\nu}$dx$^{\mu}\wedge$dx$^{\nu}$, is closed (indeed, $(g_{G,as})_{\mu\nu}$ is constant) and nondegenerate, so:
\begin{prop} The associated 2-form $\omega_G$ to $g_G$ is a symplectic form on $\R^{n+1}$.
\end{prop} 
From the nondegeneracy condition, a function on (X,$\omega$) defines an unique vector field X$_f$ by the relation: $\omega$(X$_f$,.) = - df. Consequently, one can define a bilinear operation on C($\R^{3+1}$) with a symplectic form on $\R^{3+1}$.
\begin{dfn} The \emph{Poisson bracket} $\left\{,\right\}$ associated with a symplectic form $\omega$ is defined by:
\begin{eqnarray*}
	\left\{f,g\right\} = \omega(X_f,X_g).
\end{eqnarray*}
\end{dfn}

Let's construct explicitly the bracket $\left\{,\right\}_G$ associated with $\omega_G$. We have:
\begin{eqnarray*}
	\left\{f,g\right\}_{G} &:=& g_{G,as}(X_f,X_g),\\
	&=& dx^{\mu}((X_f)dx^{\nu}(X_g) - dx^{\nu}(X_f)dx^{\mu}(X_g).	
\end{eqnarray*}
We have:
\begin{eqnarray*}
	g_{G,as}(X_f,.) &=& g^{G,as}_{\mu\nu}\left[dx^{\mu}((X_f)dx^{\nu} - dx^{\nu}(X_f)dx^{\mu}\right]\\
	&=& - df = - \partial_{\rho}fdx^{\rho}.\\
\Rightarrow \partial_{\rho}f 
&=&  g^{G,as}_{\mu\nu}\left[dx^{\mu}(X_f)\delta^{\nu}_{\rho} - dx^{\nu}(X_f)\delta^{\mu}_{\rho}\right],\\
	&=& 2g^{G,as}_{\mu\nu}dx^{\mu}(X_f)\delta^{\nu}_{\mu},\\
	&=& 2g^{G,as}_{\mu\rho}dx^{\mu}(X_f),\\
	\Rightarrow \left\{f,g\right\}_{G} &=& - \frac{1}{2}\partial_{\nu}f(g_{G,as}^{-1})^{\nu\eta}g_{\eta\tau}dx^{\tau}(X_g) + \frac{1}{2}\partial_{\nu}g(g_{G,as}^{-1})^{\nu\beta}g_{\beta\chi}dx^{\chi}(X_f),\\
	&=& \frac{1}{4}(g_{G,as}^{-1})^{\nu\eta}\partial_{\nu}f\partial_{\eta}g - \frac{1}{4}(g^{-1})^{\nu\beta}\partial_{\nu}g\partial_{\beta}f,\\
	&=& \frac{1}{4}(g^{-1})^{\nu\eta}\left[\partial_{\nu}f\partial_{\eta}g - \partial_{\nu}g\partial_{\eta}f\right].
\end{eqnarray*}
Furthermore, a simple calculation of the inverse form $(g_{G,as})^{-1}$ gives (with a slight abuse of notation):
\begin{eqnarray*}
	(g_{G,as})^{-1} = 
	\begin{pmatrix}
    \left[0_3\right]  &  - \vec{a}/|\vec{a}|^2\\
    \vec{a}/|\vec{a}|^2 &  0  
  \end{pmatrix}
  .
\end{eqnarray*}
and finally, 
\begin{eqnarray}
	\left\{f,g\right\}_{G} = \frac{\vec{a}}{4|\vec{a}|^2}\cdot (\partial_0f\vec{\partial} g - \partial_0g\vec{\partial} f).
\end{eqnarray}

Here are values of our bracket for some particular functions:
\begin{eqnarray*}
	\left\{x^{\nu},x^{\mu}\right\}_{G} &=& \frac{1}{4|\vec{a}|^2}a^i (\delta^{\mu}_0\delta^\nu_i - \delta^{\nu}_0\delta^\mu_i) \quad \textrm{so } \left\{t,x^{i}\right\}_{G} = \frac{1}{4|\vec{a}|^2}a^i, \quad \left\{x^{i},x^{j}\right\}_{G} = 0.\\
	\left\{x^{\mu},f\right\}_{G} &=& \frac{1}{4|\vec{a}|^2} \vec{a} \cdot (\delta^{\mu}_0 \vec{\partial}f - \partial_0f \vec{\partial}x^\mu) \quad \textrm{i.e. } \left\{t,f\right\}_{G} = \frac{1}{4|\vec{a}|^2}\vec{a} \cdot \vec{\partial}f , \quad \left\{x^i,f\right\}_{G} = - \frac{1}{4|\vec{a}|^2}a^i \partial_0f .
\end{eqnarray*}

The two last equalities express the fact that vector fields defined by functions t and $\vec{x}$ are (up to factor) $\vec{a}\cdot \vec{\partial}$ and $\vec{a} \partial_0$ respectively.

\section{Conclusion}
We know that Galilean relativity serves as a reliable physical law when the characteristic speed of the studied physical system is small compared to the speed of light. Therefore, it remains a valuable tool for understanding nature. In this paper, we have encountered the Galilean invariant vector ($\vec{a}$,a$^0$), the correct physical interpretation of which is still lacking. Furthermore, it would be interesting to investigate the geometry related to Galilean bilinear forms. 

\subparagraph{Acknowledgements} 
I would like to thank my supervisor Roland Raboanary, and Christian Rakotonirina for their critical comments and for carefully reading the manuscript.


\begin{thebibliography}{4}
\bibitem{raan02} Raoelina Andriambololona, \emph{Alg\`ebre Multilin\'eaire}, Lecture given at the University of Antananarivo, \textbf{2002}.
\bibitem{anca00} A. Cannas da Silva, \emph{Lectures on Symplectic Geometry}, based on course given at UC Berkeley, \textbf{2000}.
\end{thebibliography}
\end{document}